\documentclass[final,3p,times,twocolumn]{elsarticle}
\usepackage{graphics} 

\journal{JP Journal of Solids and Structures}

\begin{document}

\begin{frontmatter}



\title{Electron density distribution of bilayer nanographene\\
and band structures of boron-carbon-nitride systems}


\author{Kikuo Harigaya and Hiroshi Imamura}

\address{Nanosystem Research Institute, AIST, 
Tsukuba 305-8568, Japan}

\begin{abstract}
Bilayer graphene nanoribbon with zigzag edge is investigated with the
tight binding model. Two stacking structures, $\alpha$ and $\beta$,
are considered.  The band splitting is seen in the $\alpha$ structure,
while the splitting in the wave number direction is found
in the $\beta$ structure.  The local density of states in the
$\beta$ structure tend to avoid sites where interlayer hopping 
interactions are present.  The calculation is extended to the
boron-carbon-nitride systems.  The qualitative properties
persist when zigzag edge atoms are replaced with borons
and nitrogens.
\end{abstract}


\end{frontmatter}


\section{Introduction}

The graphite, multi-layer, and single-layer graphene materials 
have been studied intensively, since the electric field effect 
has been found in atomically thin graphene films [1]. These 
materials can be regarded as bulk systems. On the other hand, 
nanographenes with controlled edge structures have been predicted 
to have localized states along the zigzag edges [2].  
The presence of the edge states has been observed by experiments 
of scanning tunneling spectroscopy [3,4].  Thus, the 
studies of the edge states are one of the interesting topic of 
the field.  The recent atomic bottom-up fabrication of nanoribbons
really promotes experimental and theoretical investigations [5].

In the previous paper [6], the tight binding model has been 
solved numerically, and effects of interlayer interactions 
have been considered. In the $\beta$ structure [Fig. 1 (b)], 
the split of the energy bands is not seen compared 
with that of the $\alpha$ structure [Fig. 1 (a)].  
The difference of the band split in the $\alpha$ and $\beta$ 
structures appears.  Number of states in the energy window for 
the positive wave number has been calculated for the single 
layer, $\alpha$, and $\beta$ structures. The dependence of 
the energy reflects the band  structures, and this will 
appear in quantization of conductance experimentally.

In this paper, we will report the electron density
distribution in nanographene materials with zigzag 
edges including the inter-layer interactions.
The calculation is extended to the boron-carbon-nitride 
systems, too.  We report the band structures of the
single and bilayer systems.

\section{Model}

The following tight binding model is used in the calculation:
\begin{eqnarray}
H=-t \sum_{\langle i,j \rangle, \sigma} 
(c_{i,\sigma}^\dagger c_{j,\sigma} + {\rm h.c.}) \nonumber \\
-t \sum_{\langle i,j \rangle, \sigma}
(d_{i,\sigma}^\dagger d_{j,\sigma} + {\rm h.c.}) \nonumber \\
-t_1 \sum_{\langle i,j \rangle, \sigma} 
(c_{i,\sigma}^\dagger d_{j,\sigma} + {\rm h.c.}),
\end{eqnarray}
where $c_{i,\sigma}$ and $d_{i,\sigma}$ are the annihilation
operators of electrons at the lattice site $i$ of the spin $\sigma$
on the upper and lower layers, respectively.  The quantity $t$
is the hopping integral of $\pi$ electrons between neighboring
lattice sites.  Two stacking patterns,
shown in Figs. 1 (a) and (b), are considered.  They are named
as $\alpha$ and $\beta$ structures, respectively.  This 
convention has been used in the literature [7].
There are $N_z=10$ zigzag lines in upper and lower layers.
Inclusion of the bond alternations [8,9] would be interesting,
but it is not treated in this paper.

\noindent
\includegraphics[width=70mm]{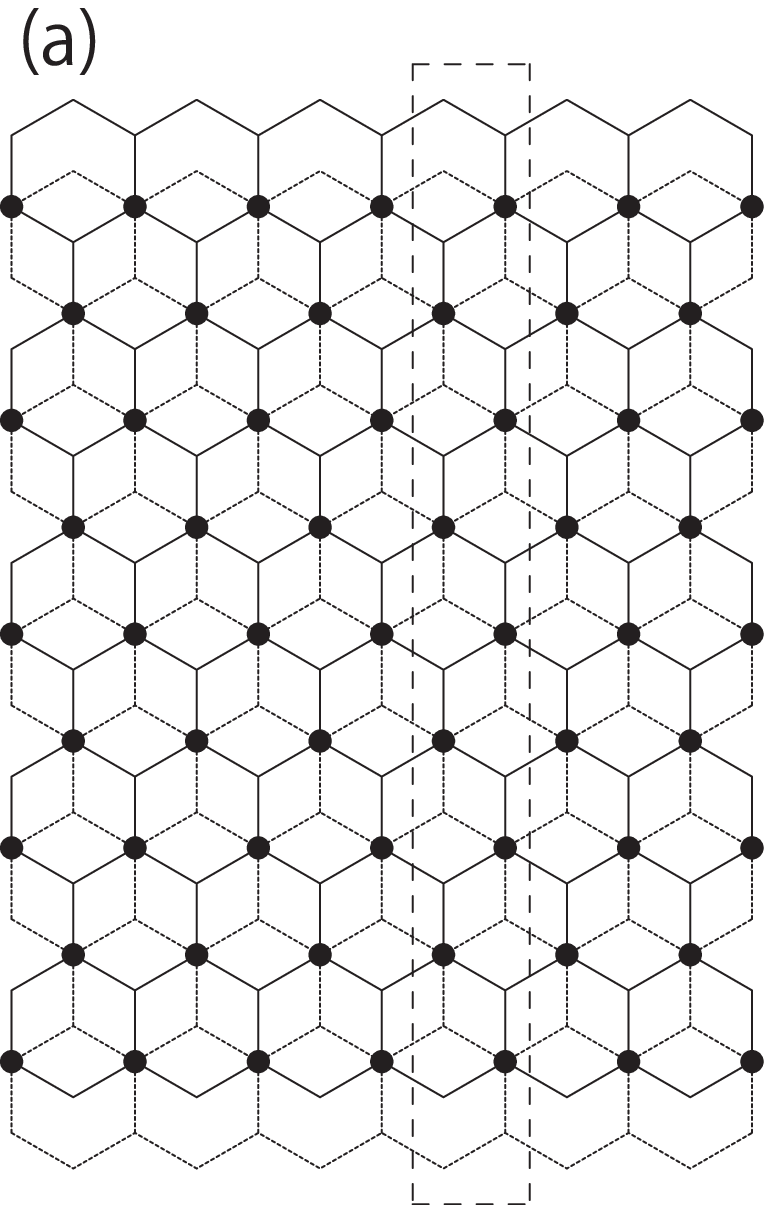}\\
\includegraphics[width=70mm]{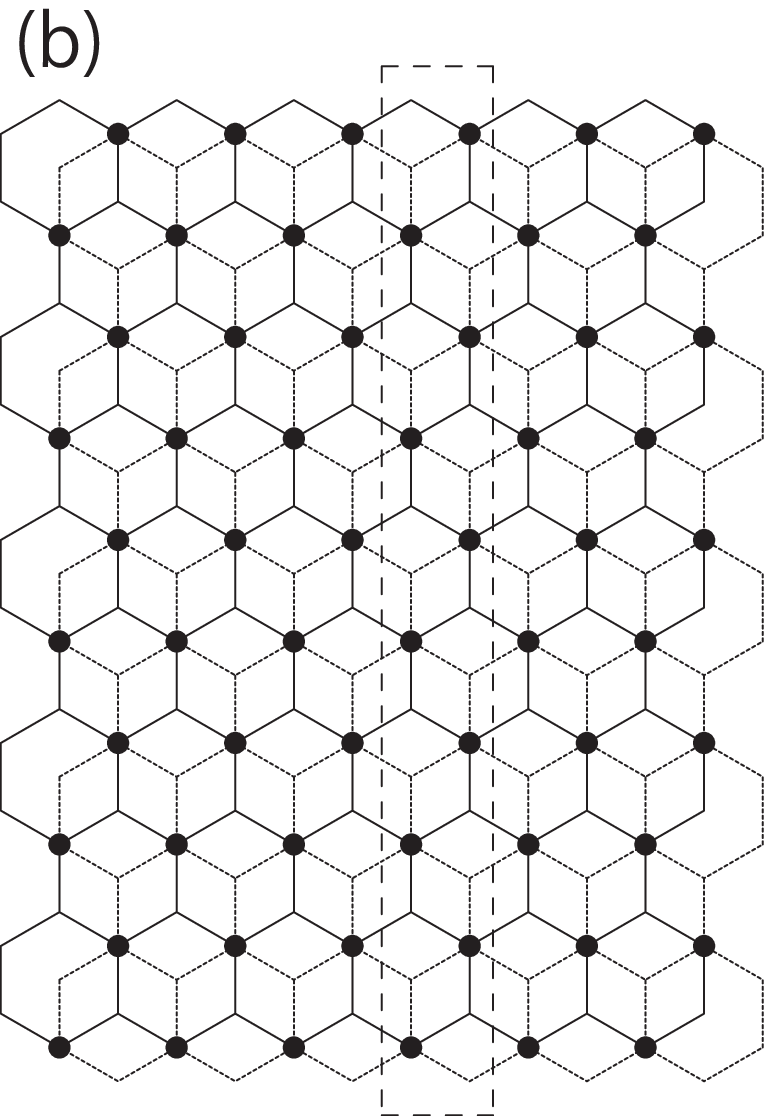}

\noindent
Fig. 1.  A-B stacked bilayer graphene nanoribbon with zigzag edges.
The upper layer is shown by the solid lines, and the lower layer 
by dotted lines.  In the $\alpha$ structure (a), the upper layer 
is shift by the bond length downward to the position of the 
lower layer.  The region surrounded by the dashed line is
the unit cell in the direction of the one dimensional
direction.  At the circles, two carbon atoms of the upper
and lower layers overlap completely, and there is the
weak hopping interaction $t_1$ here.  In (b), the $\beta$
structure is shown, where the lower layer is shift to the
right-down direction so the stacking pattern is different
from that of the $\alpha$ structure. The convention of
the $\alpha$ and $\beta$ structures is the same used in the
paper [7].  In the area surrounded by the dashed lines, 
the unit cell is displayed.

\section{Electron density in the real space}

In order to understand effects of the band split clearly,
it is useful to watch the distribution of electron density
in the real space.  Here, we calculate electron density
by changing the energy window.  Figure 2 shows 
local density of states of electrons in the unit cell
of the upper layer.   The energy is shown below each 
figure, with the number of states $N_s$ crossing at 
the energy for the region of positive wave number.
The diameter of the circle is proportional to the
electron density.  The total electron number in the
unit cell is normalized to be constant divided by $N_s$,
in order to eliminate the effects of enlargement
of circles due to large degeneracy.  In Fig. 2,
the electron density is displayed for $N_z=10$,
so there are 20 atoms in the unit cell.
In the single layer (a), the electron density is
symmetric with respect to the center of the unit
cell.  At the energy $E=0.01t$, the electron density is
large at A type sites in the upper part, while it is
large at B type sites in the lower part.  There will be
an edge state at $E=0$, but the electron density
has already intrude into the inner part at this
energy.  As the energy increases from $E=0.1t$
to $0.7t$, the electron density moves gradually
from A type to B type sites (or, from B type to A 
type sites).  After interlayer
interaction $t_1=0.1t$ is taken into account,
the spatial symmetry is lost as shown for
the $\alpha$ (b) and $\beta$ (c) structures.
The energy is changed from $E=0.01t$ to $0.75t$
in (b), and from $E=0.01t$ to $0.7t$ in (c).
$N_s$ is increased by steps of jumps.
Both in (b) and (c), the electron density is relatively
larger at odd number sites from the top of the
unit cell.  It is smaller at even number sites
in the inner part.  The result looks similar
at a glance.  However, there is a difference.
The interlayer interaction $t_1$ is present
at odd number sites of the unit cell in the upper
layer in Fig 1. (a).  So, this interaction gives
rise bonding anti-bonding split of energy band.
The electrons favor to have large amplitudes
at odd number sites in the inner part.
On the other hand, the interaction $t_1$ is
located at even number sites in Fig. 1 (b).
The electrons tend to avoid the interaction $t_1$,
and the removal of the degeneracy occurs in the
direction of the wave number.  In this way,
the electron density in the real space is
related mutually with the difference in changes,
seen in the band structures of the bilayer
nanoribbons.  These findings will be observed
possibly by conduction and STM experiments.

\noindent
\includegraphics[width=70mm]{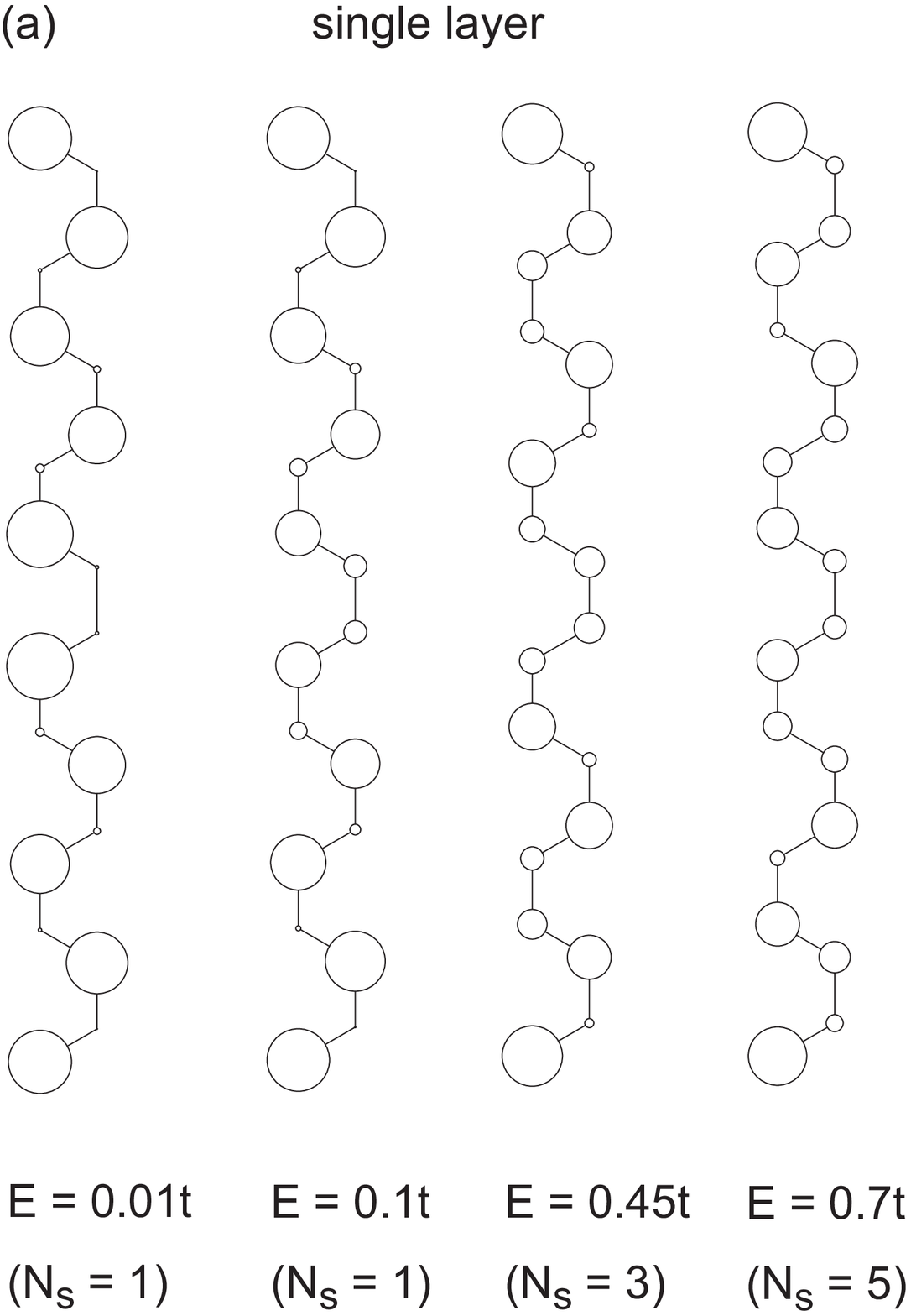}\\
\includegraphics[width=70mm]{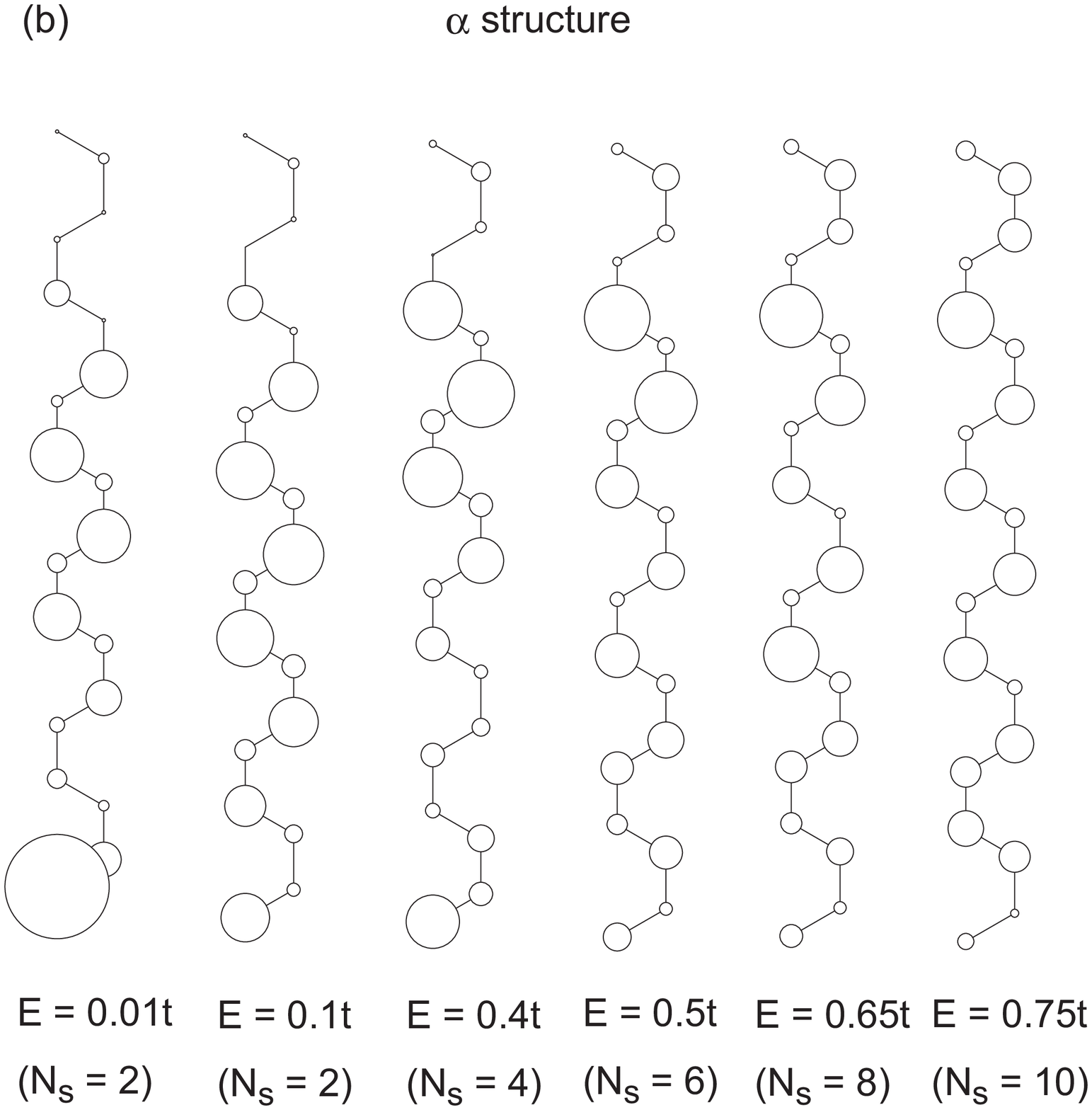}\\
\includegraphics[width=70mm]{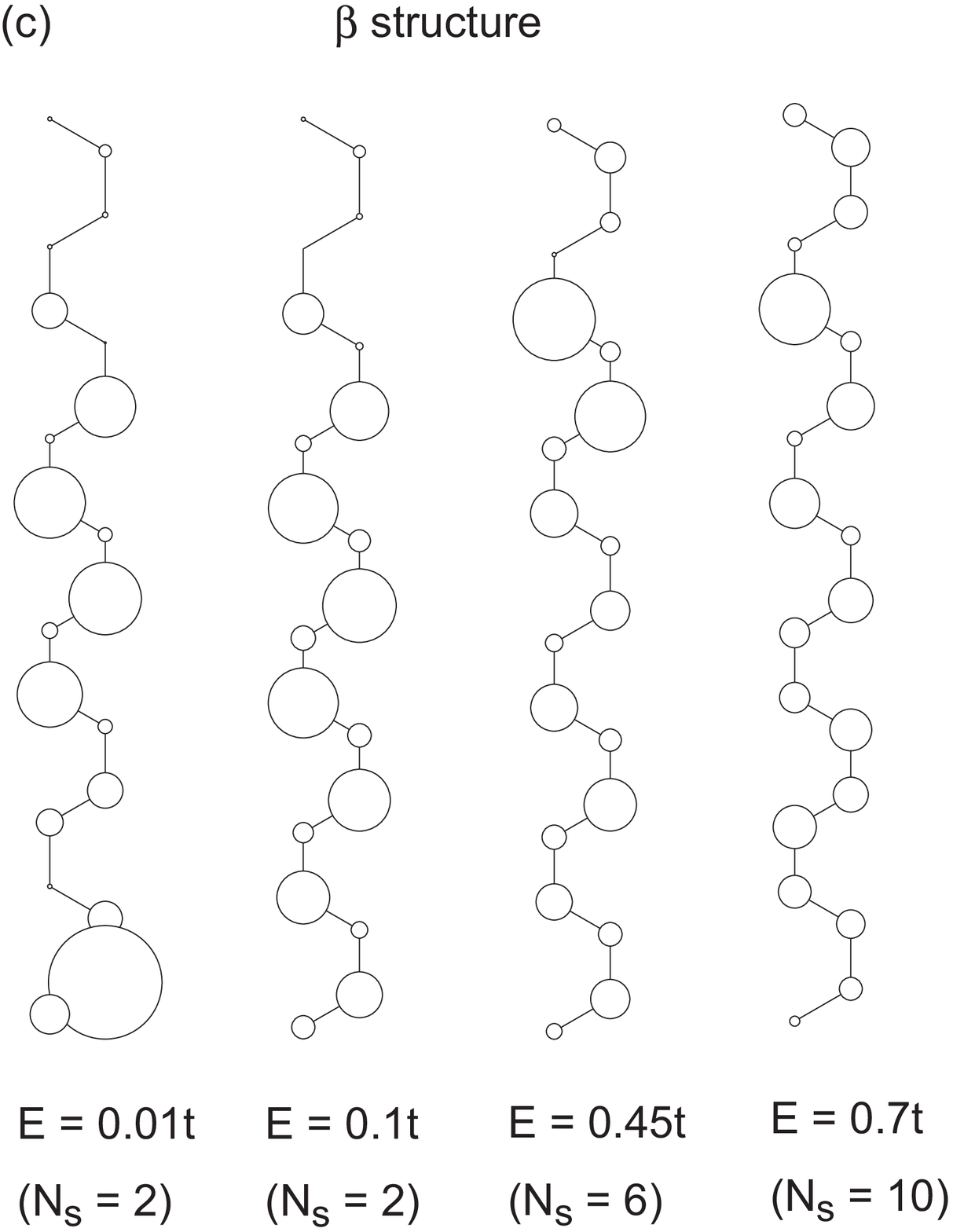}

\noindent
Fig. 2.  Local density of states in the unit cell of the upper
layer for the single layer (a), $\alpha$ structure (b),
and $\beta$ structure (c).  The density is proportional
to the diameter of the circle.  The energy is shown
below each figure, with the number of states $N_s$
crossing at the energy for the region of positive wave number.

\section{Band structures of the BCN systems}

The calculation is extended to the boron-carbon-nitride
(BCN) systems [10].  The stacking patterns are considered
with the spatial inversion symmetry.  The presence of B
and N atoms is taken into account by the site energies
E$_{\rm B}=+t$ and E$_N=-t$, as has been used in the 
paper [11].

Energy band structures of the $N_z=10$ systems
are displayed in Fig. 3, for the single layer (a), 
$\alpha$ structure (b), and $\beta$ structure (c).
The interlayer interaction strength is $t_1=0.1t$.
In (b), the split of the energy bands is seen
compared with Fig. (a).  In contrast, split in the
perpendicular direction is small in Fig. (c),
except for the lowest unoccupied state and
highest occupied state, which have large amplitude
near edge atoms. In order to see split structures clearly,
details near the Brillouin zone edge are magnified
 for the single layer (d), $\alpha$ structure (e), 
and $\beta$ structure (f).  In (e) and (f), the 
energy bands of the single layer are shown by the 
red lines for comparison.  We find that the qualitative 
properties between the difference of the energy
band splits of the $\alpha$ and $\beta$ structures
persist when zigzag edge atoms are replaced with borons
and nitrogens.  This property is confirmed by 
looking at the numerical data, also.

\noindent
\includegraphics[width=70mm]{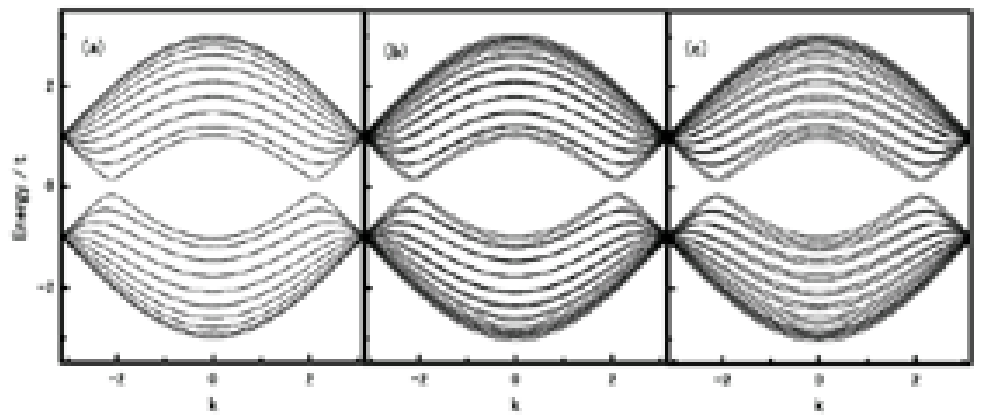}\\
\includegraphics[width=70mm]{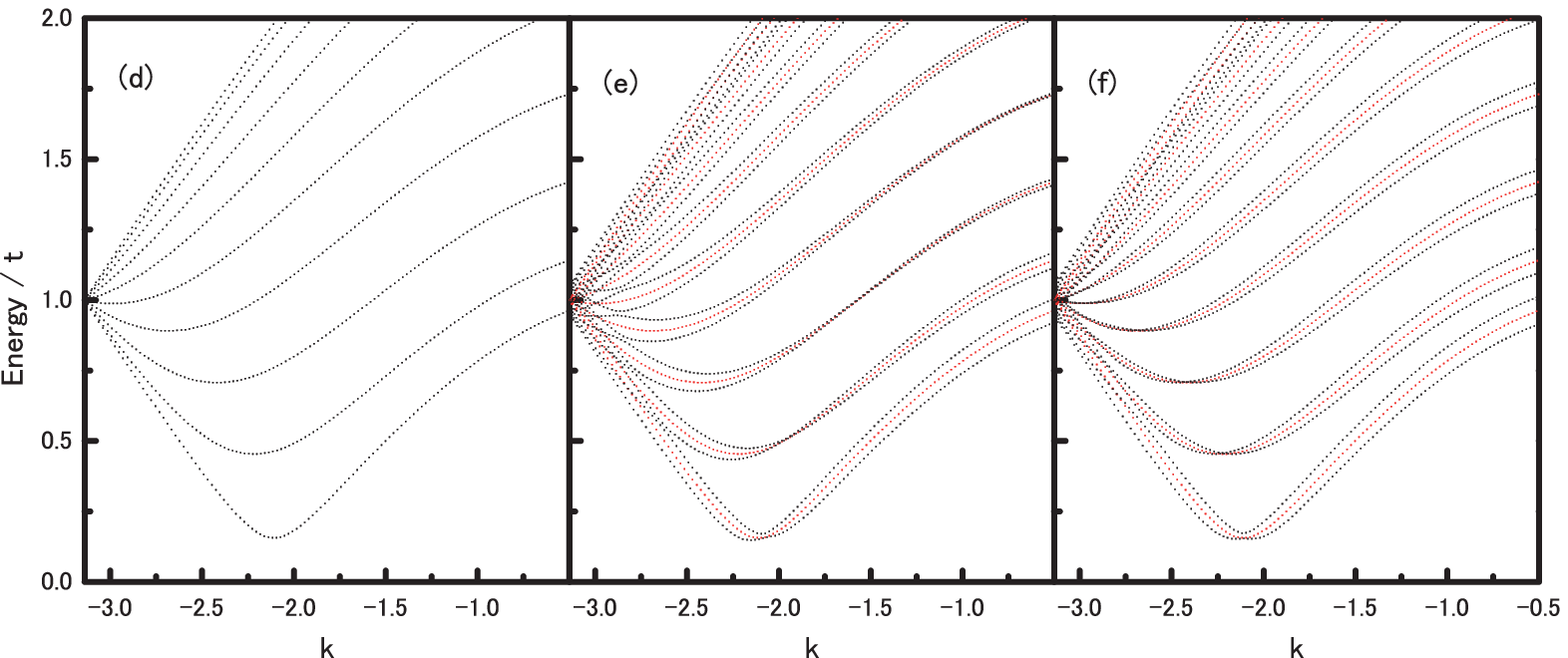}

\noindent
Fig. 3. Energy band structures of the boron-carbon-nitride
nanoribbons for the single layer (a), 
$\alpha$ structure (b), and $\beta$ structure (c).
Details near the Brillouin zone edge are magnified
 for the single layer (d), 
$\alpha$ structure (e), and $\beta$ structure (f).
In (e) and (f), the energy bands of the single
layer are shown by the red lines for comparison.

\section{Summary}

In summary, weak interlayer interactions have been considered
for the bilayer graphene nanoribbon with zigzag edge.
The $\alpha$ and $\beta$ stacking structures have been considered.  
The band splitting is seen in the $\alpha$ structure, while the 
splitting in the wave number direction is found in the $\beta$ 
structure [6].  The local density of states in the $\beta$ structure 
tend to avoid sites where interlayer hopping interactions exist.
The calculation has been extended to the boron-carbon-nitride 
systems.  The qualitative properties persist when zigzag edge 
atoms are replaced with borons and nitrogens.

\begin{flushleft}
{\bf Acknowledgemwnts}
\end{flushleft}

This research was supported by the International 
Joint Work Program of Daeduck Innopolis under the 
Ministry of Knowledge Economic (MKE) of the
Korean Government.

\begin{flushleft}
{\bf References}
\end{flushleft}

\noindent
$[1]$ K. S. Novoselov, A. K. Geim, S. V. Morozov, D. Jiang, 
Y. Zhang, S. V. Dubonos, I. V. Grigorieva, and A. A. Firsov,
Science {\bf 306}, 666 (2004).\\
$[2]$ M. Fujita, K. Wakabayashi, K. Nakada, and K. Kusakabe,
J. Phys. Soc. Jpn. {\bf 65}, 1920 (1996).\\
$[3]$ Y. Kobayashi, K. Fukui, T. Enoki, and K. Kusakabe,
Phys. Rev. B {\bf 73}, 125415 (2006).\\
$[4]$ Y. Niimi, T. Matsui, H. Kambara, K. Tagami, M. Tsukada, 
and H. Fukuyama, Phys. Rev. B {\bf 73}, 085421 (2006).\\
$[5]$ J. Cai et al., Nature {\bf 466}, 470 (2010).\\
$[6]$ K. Harigaya, H. Imamura, K. Wakabayashi,
and O. Ozsoy, Journal of Superconductivity and Novel Magnetism,
(to be published); http://arxiv.org/abs/1102.5473.\\
$[7]$ M. P. Lima, A. J. R. da Silva, and A. Fazzio,
Phys. Rev. B {\bf 81}, 045430 (2010).\\
$[8]$ O. \"{O}zsoy, J. Optoelectron. Adv. Mater. 
{\bf 9}, 2283 (2007).\\
$[9]$ O. \"{O}zsoy and K. Harigaya, J. Comput. Theor. 
Nanosci, {\bf 8}, 31 (2011).\\
$[10]$ K. Harigaya, Jpn. J. Appl. Phys. {\bf 45}, 
7237 (2006).\\
$[11]$ T. Yoshioka, H. Suzuura, and T. Ando,
J. Phys. Soc. Jpn. {\bf 72}, 2656 (2003).\\ 

\end{document}